\documentclass[aps,twocolumn,showpacs,showkeys,preprintnumbers,amsmath,amssymb]{revtex4} 
\usepackage[]{graphicx}
\usepackage{amsmath}

\def\intii{\int_{-\infty}^{\infty}}

\begin{document} 

\title{Length scale competition in soliton-bearing systems: \\
A collective coordinate approach}

\author{Sara Cuenda}
\email{scuenda@math.uc3m.es}

\author{Angel S\'anchez$^{*,}$}
\email{anxo@math.uc3m.es}

\affiliation{%
$^*$Grupo Interdisciplinar de Sistemas Complejos (GISC) and
Departamento de Matem\'aticas\\
Universidad Carlos III de Madrid, Avenida de la Universidad 30, 28911
Legan\'es, Madrid, Spain}

\homepage{http://gisc.uc3m.es}

\affiliation{%
$^{\dag}$Instituto de Biocomputaci\'on y  F\'{\i}sica de
Sistemas Complejos (BIFI),\\
Facultad de Ciencias, Universidad de Zaragoza, 50009 Zaragoza, Spain
}%

\date{\today}% It is always \today, today,

\begin{abstract}
We study the phenomenon of length scale competition, an instability of 
solitons and other coherent structures that takes place when their size is of
the same order of some characteristic scale of the system in which they propagate.
Working on the framework of nonlinear Klein-Gordon models as a paradigmatic 
example, we show that this instability can be understood by means of a collective
coordinate approach in terms of soliton position and width. As a consequence, 
we provide a quantitative, natural explanation of the phenomenon in much
simpler terms than any previous treatment of the problem. Our technique 
allows to study the existence of length scale competition in most soliton 
bearing nonlinear models and can be extended to coherent structures with 
more degrees of freedom, such as breathers. 
\end{abstract}

\keywords{Solitons, Coherent structures, 
Space-dependent perturbations, Length-scale competition, Collective coordinates}

\pacs{05.45.Yv, 03.75.Lm, 02.30.Jr}

\maketitle

\noindent{\bf LEAD PARAGRAPH}\\

{\bf 
Solitons, solitary waves, vortices and other coherent structures possess, 
generally speaking, a characteristic length or size. One important feature
of these coherent structures, which usually are exact solutions of 
certain nonlinear models, is their robustness when the corresponding models are 
perturbed in different ways. A case relevant in many real applications is 
that of space-dependent perturbations, that may or may not have their own 
typical length scale. Interestingly, in the latter case, it has been known
for a decade that when the perturbation length scale is comparable to the
size of the coherent structures, the effects of even very small perturbing
terms are dramatically enhanced. Although some analytical approaches have
shed some light of the mechanisms for this special instability, a clear-cut,
simple explanation was lacking. In this paper, we show how such explanation 
arises by means of a reduction of degrees of freedom through the so-called
collective coordinate technique. The analytical results have a straightforward
physical interpretation in terms of a resonant-like phenomenon. 
Notwithstanding the fact that we work on 
a specific class of soliton-bearing equations, our approach is readily 
generalizable to other equations and/or types of coherent structures. 
}

\newpage

\section{Introduction}

Fifty years after the pioneering discoveries of Fermi, Pasta and Ulam 
\cite{FPU}, the paradigm of coherent structures has proven itself one of
the most fruitful ones of Nonlinear Science \cite{Scott}. Fronts, 
solitons, solitary waves, breathers, or vortices are instances of such
coherent structures of relevance in a plethora of applications in 
very different fields. One of the chief reasons that gives all these
nonlinear excitations their paradigmatic character is their robustness
and stability: Generally speaking, when systems supporting these 
structures are perturbed, the structures continue to exist, albeit with 
modifications in their parameters or small changes in shape
(see \cite{YuriRMP,IJMPB} for reviews). This 
property that all these objects (approximately) retain their identity 
allows one to rely on them to interpret the effects of perturbations 
on general solutions of the corresponding models. 

Among the different types of coherent structures one can encounter, 
topological solitons are particularly robust due to the existence of 
a conserved quantity named topological charge. Objects in this class 
are, e.g., kinks or vortices and can be found in systems ranging from
Josephson superconducting devices to fluid dynamics. A particularly 
important representative of models supporting topological solitons 
is the family of nonlinear Klein-Gordon equations \cite{Scott}, 
whose expression is 
\begin{equation}\label{kg}
\phi_{tt}-\phi_{xx}+\frac{dU}{d\phi}=0.
\end{equation}
Specially important cases of this equation occur when 
$U(\phi)=\frac{1}{4}(\phi^2-1)^2$, giving the so-called $\phi^4$ 
equation, and when 
$U(\phi)=1-\cos(\phi)$, leading to the sine-Gordon (sG) equation,
which is one of the few examples of fully integrable systems
\cite{AKNS}. 
Indeed, for any initial data the solution of the sine-Gordon  
equation can be expressed as a sum of kinks (and antikinks), 
breathers, and linear waves. Here we focus on kink solitons, 
which have the form 
\begin{equation}\label{kink_sg}
\phi(x,t)=4\arctan\left\{\exp\left(\frac{x-vt}{\sqrt{1-v^2}}\right)\right\},
\end{equation}
$0\leq v<1$ being a free parameter that specifies the kink velocity. 
The topological character of these solutions arises from the fact that they
join two minima of the potential $U(\phi)$, and therefore they cannot be 
destroyed in an infinite system. 
Our other example, the $\phi^4$ equation, is not integrable, but supports
topological, kink-like solutions as well, given by 
\begin{equation}\label{kink_f4}
\phi(x,t)=\tanh\left(\frac{x-vt}{\sqrt{2(1-v^2)}}\right).
\end{equation}

It is by now well established, already from pioneering works in the 
seventies \cite{Scott-McLaughlin,Bishop} that both types of kinks behave, 
under a wide class of perturbations, like relativistic particles. 
The relativistic character arises from the Lorentz invariance of their 
dynamics, see Eq.\ (\ref{kg}), and implies that there is a maximum 
propagation velocity for kinks (1 in our units) and their characteristic width 
decreases with velocity. Indeed, even behaving as particles, kinks do have
a characteristic width; however, for most perturbations, that is not a 
relevant parameter and one can consider kinks as point-like particles. 
This is not the case when the perturbation itself gives rise to certain
length scale of its own, a situation that leads to the phenomenon of 
length-scale competition, first reported in \cite{yo1} 
(see \cite{SIAM} for a review). This phenomenon is nothing but 
an instability that occurs when the length of a coherent structure
approximately matches that of the perturbation: Then, small values of
the perturbation amplitude are enough to cause large modifications or
even destroy the structure. Thus, in \cite{yo1}, the perturbation 
considered was sinusoidal, of the form
\begin{equation}
\phi_{tt}-\phi_{xx}+\frac{dU}{d\phi}(1+\epsilon\cos(kx))=0,
\end{equation}
where $\epsilon$ and $k$ are arbitrary parameters. The structures studied here 
were breathers, which are exact solutions of the sine-Gordon equation with a 
time dependent, oscillatory mode (hence the name `breather') and that can be
seen as a bound kink-antikink pair. It was found that small 
$k$ values, i.e., long perturbation wavelengths, induced breathers to move
as particles in the sinusoidal potential, whereas large $k$ or equivalent
short perturbation wavelengths, were unnoticed by the breathers. In the 
intermediate regime, where length scales were comparable, breathers 
(which are non topological) were destroyed.  

As breathers are quite complicated objects, the issue of length scale 
competition was addressed for kinks in \cite{yo2}. In this case, kinks
were not destroyed because of the conservation of the topological 
charge, but length scale competition was present in a different way:
Keeping all other parameters of the equation constant, it was observed
that kinks could not propagate when the perturbation wavelength was of
the order of their width. In all other (smaller or larger) perturbations,
propagation was possible and, once again, easily understood in terms of 
an effective point-like particle. Although a explanation of this 
phenomenon was provided in \cite{yo2} in terms of a (numerical) linear
stability analysis and the radiation emitted by the kink, it was not a 
fully satisfactory argument for two reasons: First, the role of the 
kink width was not at all transparent, and second, there were no simple 
analytical results. These are important issues because length scale 
competition is a rather general phenomenon: It has been observed in 
different models (such as the nonlinear Schr\"odinger equation \cite{rainer})
or with other perturbations, including random ones \cite{garnier}. 
Therefore, having a simple, clear explanation of length scale competition
will be immediately of use in those other contexts. 

The aim of the present paper is to show that length scale competition can 
be understood through a collective coordinate approximation. 
Collective coordinate approaches were introduced in \cite{Scott-McLaughlin,Bishop}
to describe kinks as particles (see \cite{Scott,YuriRMP,IJMPB,SIAM} for a 
very large number of different techniques and applications of this idea). 
Although the original approximation was to reduce the equation of motion for
the kink to an ordinary differential equation for a time dependent, collective
coordinate which was identified with its center, it is being realized lately
that other collective coordinates can be used instead of or in addition to 
the kink center. One of the most natural additional coordinates to consider
is the kink width, an approach that has already produced new and unexpected
results such as the existence of anomalous resonances \cite{niurka1} or
the rectification of ac drivings \cite{luis}. There are also cases in which 
one has to consider three or more collective coordinates (see, e.g., 
\cite{matthias}). It is only natural then to apply these extended collective
coordinate approximations to the problem of length scale competition, in 
search for the analytical explanation needed. As we will see below, 
taking into account the kink width dependence on time is indeed enough 
to reproduce the phenomenology observed in the numerical simulations. 
Our approach is detailed in the next Section, whereas in Sec.\ 3 we 
collect our results and discuss our conclusions.

\section{Collective coordinate approach}

We now present our collective coordinate approach to the problem of 
length scale competition for kinks \cite{yo2}. We will use the 
lagrangian based approach developed in \cite{elias}, which is very
simple and direct. Equivalent results can be obtained with the so-called
generalized travelling wave {\em Ansatz} \cite{mertens}, somewhat more
involved in terms of computation but valid even for systems that cannot
be described in terms of a lagrangian. 

Let us consider the
generically perturbed Klein-Gordon
equation:
\begin{equation}\label{kg_pert}
\phi_{tt}-\phi_{xx}+\frac{dU}{d\phi}+\epsilon f(x,t)g(\phi)=0. 
\end{equation}
The starting point of our approach
is the lagrangian for the above equation, which is given by
\begin{eqnarray}
L&=&\intii dx\left\{\frac{1}{2}\dot{\phi}^2-\frac{1}{2}\phi_x^2-U(\phi)\right.
\nonumber\\
& &+\left.\epsilon g(\phi)\phi_x\int_{x_0}^x dy\,f(y,t)\right\}.\label{lag_pert}
\end{eqnarray}

As stated above, we now focus on the 
behaviour of kink excitations of the form \eqref{kink_f4} and \eqref{kink_sg}.
To do so, we will use a two collective coordinates approach
by substituting the \emph{Ansatz}
\begin{equation}\label{ansatz_f4}
\phi(z(t))=\tanh\left(z(t)\right)
\end{equation}
in the lagrangian of the $\phi^4$ system, and 
\begin{equation}\label{ansatz_sg}
\phi(z(t))=4\arctan\left\{\exp\left(z(t)\right)\right\}
\end{equation}
in $sG$, where $z(t)=\frac{x-X(t)}{l(t)}$ and $X(t)$ and $l(t)$ are two
collective coordinates that represent the position of the center and the
width of the kink, respectively. Substituting the expresions \eqref{ansatz_f4}
and \eqref{ansatz_sg} in the lagrangian \eqref{lag_pert} with 
our perturbation,
$f(x,t)=\cos (kx)$ and $g(\phi)=\frac{dU}{d\phi}$, we obtain an
expresion of the lagrangian in terms of $X$ and $l$,
\begin{eqnarray}
L&=&\frac{M_0l_0}{2l}\dot{X}^2+\frac{\alpha M_0l_0}{2l}\dot{l}^2-
\frac{M_0}{2}\left(\frac{l_0}{l}+\frac{l}{l_0}\right)\nonumber\\
& &-\frac{\epsilon}{k}\cos(kX)w(a)\vert_{a=kl},
\end{eqnarray}
where, for the $\phi^4$ system, $M_0=4/(3\sqrt{2})$, $l_0=\sqrt{2}$ and
$\alpha=(\pi^2-6)/12$, and for the sG system, $M_0=8$, $l_0=1$ and
$\alpha=\pi^2/12$, and
\begin{equation}
w(a)=\intii dz\,\tanh(z)(\phi '(z))^2\sin(az),
\end{equation}
which is 
\begin{equation}
w(a)=\frac{\pi a^2(a^2+4)}{24\sinh\left(\frac{\pi a}{2}\right)}
\end{equation}
for $\phi^4$ and
\begin{equation}
w(a)=\frac{2\pi a^2}{\sinh\left(\frac{\pi a}{2}\right)}
\end{equation}
for the sG model. We note that the effect of a spatially periodic perturbation
like the one considered here was studied for the $\phi^4$ model in \cite{phi4},
although the authors were not aware of the existence of length scale competition
in this system and focused on unrelated issues. 

The equations of motion of $X$ and $l$ can now be obtained using the
Lagrange equations,
\begin{equation}
\frac{d}{dt}\left(\frac{\partial L}{\partial \dot{Y}_i}\right)=
\frac{\partial L}{\partial Y_i},
\end{equation}
where $Y_i$ stands for the collective coordinates $X$ and $l$.
The ODE system for $X$ and $l$ is, finally,
\begin{eqnarray}
\dot{P}&=&\epsilon\sin(kX)w(a)\vert_{a=kl}, \\
\dot{Q}&=&-\frac{1}{2M_0l_0}\left(P^2+\frac{Q^2}{\alpha}\right)+
\frac{M_0l_0}{2}\left(\frac{1}{l^2}-\frac{1}{l_0^2}\right)\nonumber\\
& &-\epsilon\cos(kX)w'(a)\vert_{a=kl},
\end{eqnarray}
where $P=M_0l_0\dot{X}/l$ and $Q=\alpha M_0l_0\dot{l}/l$.

\begin{figure}
\begin{center}
\includegraphics[width=6cm, angle=270]{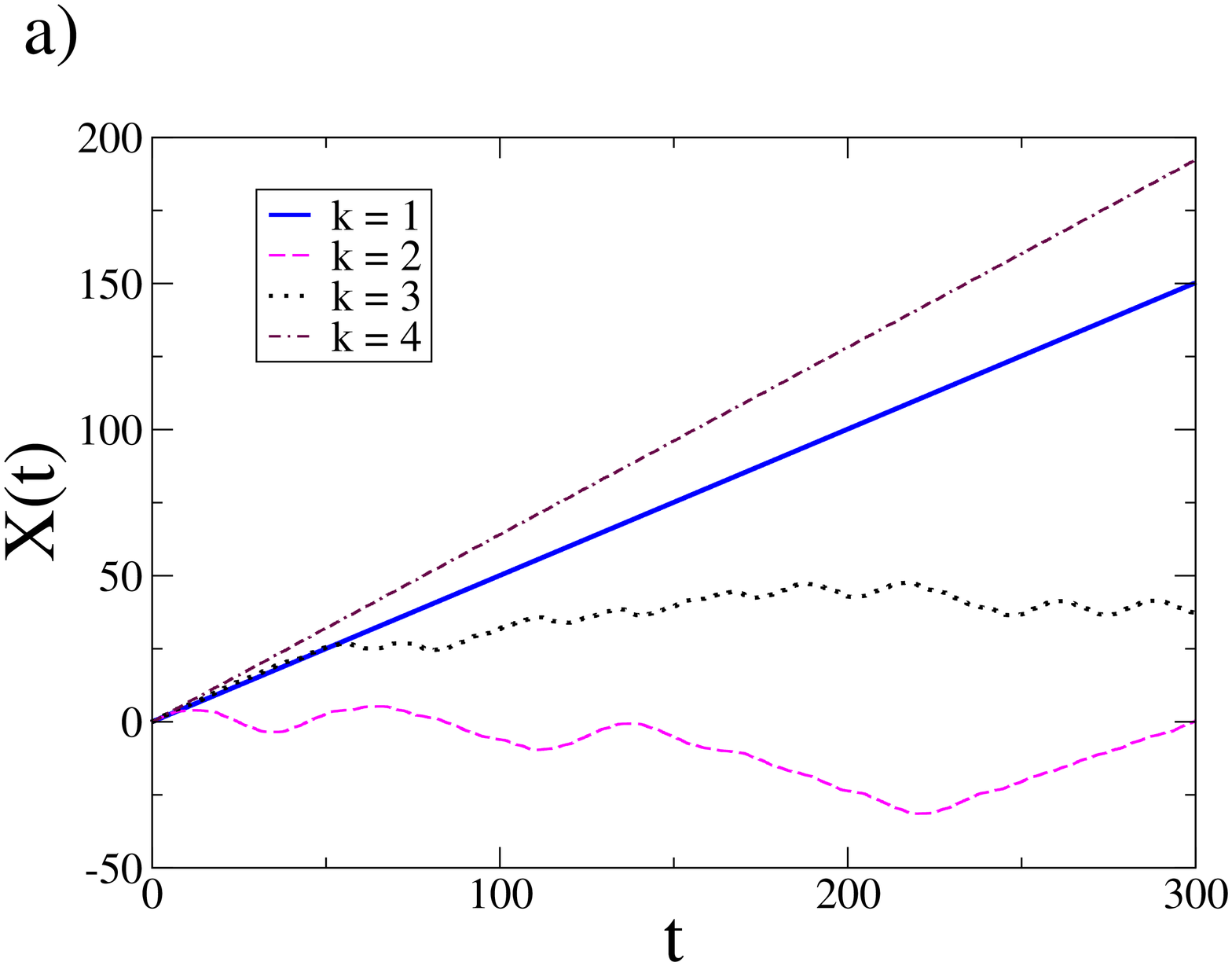}
\includegraphics[width=6cm, angle=270]{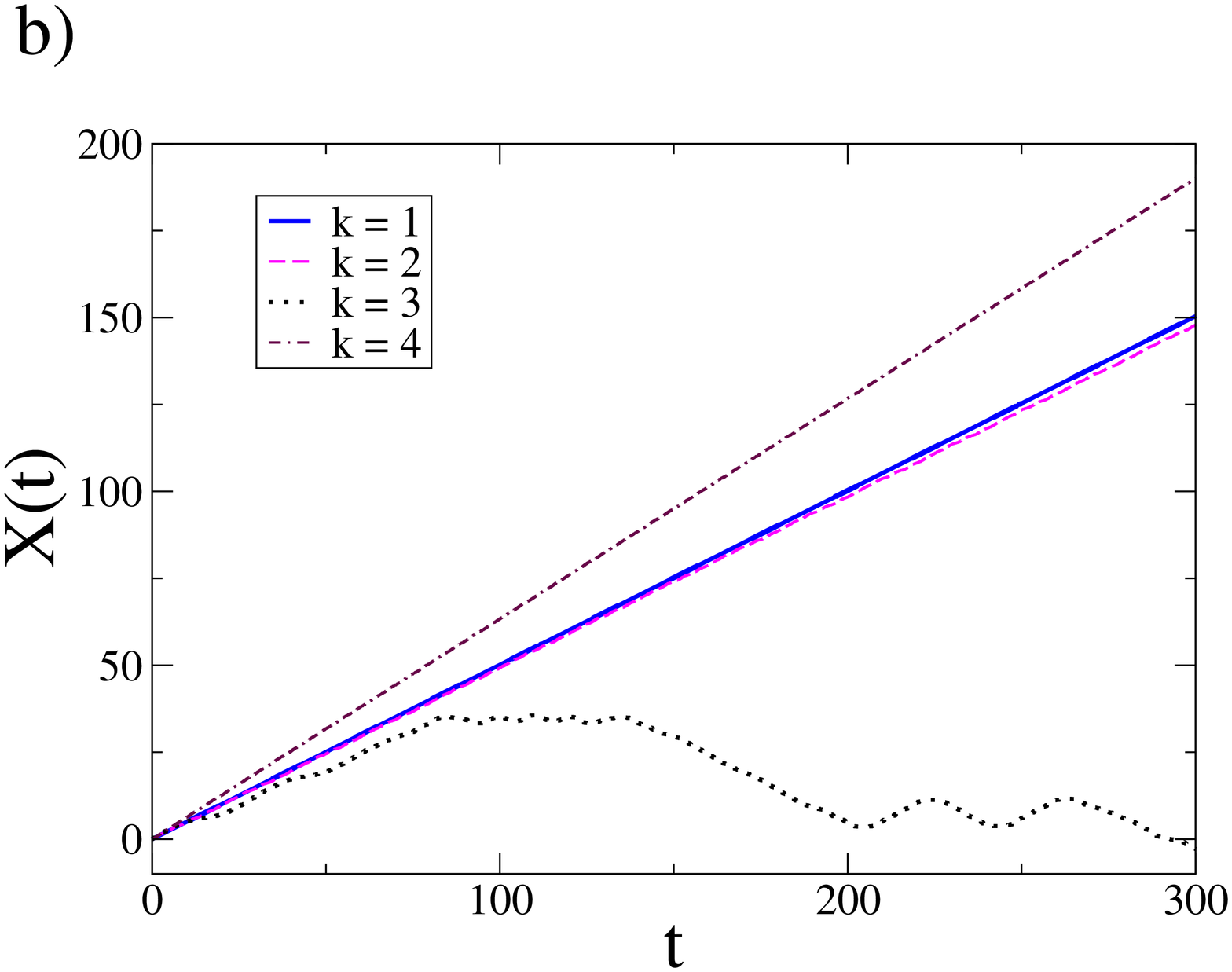}
\caption{\label{fig:lengthscale}
Different behaviours of $X(t)$ for different values of $k$ for $\epsilon=0.7$,
$X(0)=0$, $\dot{X}(0)=0.5$, $l(0)=l_0(1-\dot{X}(0)^2)^{1/2}$ and $\dot{l}(0)=0$
in the a) $\phi^4$ model, b) sG model.
}
\end{center}
\end{figure}

\begin{figure}
\begin{center}
\includegraphics[width=7cm, angle=270]{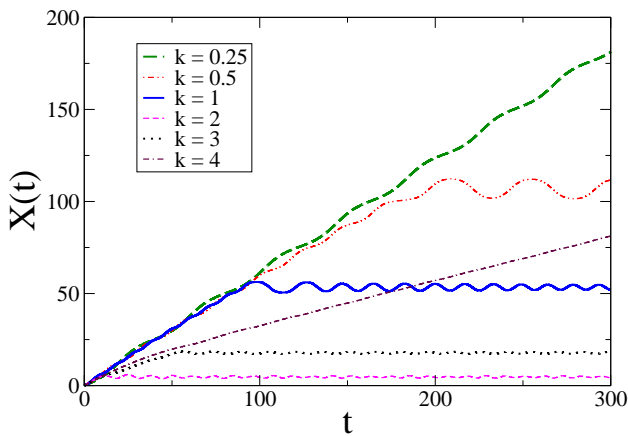}
\caption{\label{fig:simphi4}
Different behaviours of $X(t)$ for different values of $k$ for $\epsilon=0.7$,
$X(0)=0$, $\dot{X}(0)=0.5$ as obtained from numerical simulations 
of the full $\phi^4$ model.
}
\end{center}
\end{figure}

The equations above are our final result for the dynamics of sG and 
$\phi^4$ kinks in terms of their center and width. As can be seen, they
are quite complicated equations and we have not been able to solve them
analytically. Therefore, in order to check whether or not they predict 
the appearance of length scale competition, we have integrated them 
numerically using a Runge-Kutta scheme \cite{recipes}.
The simulation results for different values of $k$, $X(0)=0$ 
are in Fig. \ref{fig:lengthscale}. We have taken 
$\epsilon=0.7$  in order to compare with the numerical results for the
sG equation in \cite{yo2}. The plots present already the physical 
variable $X(t)$, i.e., the position of the center of the kink. As
we may see, the agreement with the results in \cite{yo2} is 
excellent for the sG equation (cf.\ Figs.\ 7 and 8 in \cite{yo2}):
For small and large wavelengths the kink, initially at the top of one
of the maxima of the perturbation, travels freely; however, for 
intermediate scales, the kink is trapped and even moves backward
for a while. As previous works \cite{phi4} on the same problem for 
the $\phi^4$ equation did not touch upon this issue, we have carried 
out numerical simulations, also using a Runge-Kutta scheme,
of the full partial differential equation.
The results, presented in Fig.\ \ref{fig:simphi4}, confirm once again the
agreement with the prediction of the collective coordinate approach. 
We see that for $k=2$, the value at which the collective coordinate 
prediction is more dramatic, the kink is trapped at the first potential 
well, indicating that for that value the length scale competition 
phenomenon is close to its maximum effect. 
Interestingly, a more detailed analysis of Fig.\ \ref{fig:simphi4} shows 
two different kink trapping processes. 
In agreement with the collective coordinate prediction, for $k=2,3$ we see
that the kink is trapped very early, after having travelled a few potential
wells at most. This is the length scale competition regime. However, there 
is an additional trapping mechanism: As is well known \cite{phi4}, kinks 
travelling on a periodic potential emit radiation. This process leads to 
a gradual slowing down of the kink until it is finally unable to proceed 
over a further potential well. This takes place at a much slower rate than 
length scale competition and hence the kink is stopped after travelling 
a larger distance in the system. The radiation emission as observed in 
the simulation is smoother than in the length scale competition trapping. 

\section{Discussion and conclusions}

As we have seen in the preceding section, a collective coordinate approach 
in terms of the kink center and width is able to explain in a correct 
quantitative manner the phenomenon 
of length scale competition, observed in numerical simulations earlier 
for the sG equation with a spatially periodic perturbation \cite{yo2,SIAM}.
The structure of the equations makes it clear the necessity for a second 
collective coordinate; imposing $l(t)=l_0$ constant, we recover the equation
for the center already derived in \cite{yo2}, which shows no sign at all of 
length scale competition, predicting effective particle-like behavior for
all $k$. The validity of this approach has been also shown in the context
of the $\phi^4$ equation, which had not been considered before from this
viewpoint. In spite of the fact that the collective coordinate equations 
cannot be solved analytically, they provide us with the physical explanation
of the phenomenon in so far as they reveal the key role played by the 
width changes with time and their coupling with the translational degree 
of freedom. 

It is interesting to reconsider the analysis carried out in \cite{yo2} of
length scale competition through a numerical linear stability analysis. 
In that previous work, it was argued that the instability arose because,
for the relevant perturbation wavelengths, radiation modes moved 
below the lowest phonon band, inducing the emission of long wavelength 
radiation which in turn led to the trapping of the kink. It was also 
argued that those modes became internal modes, i.e., kink shape deformation
modes in the process. The approach presented here is a much more simple 
way to account for these phonon effects: Indeed, as was shown by Quintero 
and Kevrekidis \cite{niurka2}, (odd) phonons do give
rise to width oscillations very similar to those induced by an internal
mode \cite{niurka2}. We are confident that what our perturbation 
technique is making clear is precisely the
result of the action of those phonons, summarized in our approach in
the width variable $l(t)$. The case for the $\phi^4$ equation is sligthly
different: Whereas the sG kink does not have an internal mode \cite{niurka3}
and $l(t)$ is hence a collective description of phonon modes, the $\phi^4$
kink possesses an intrinsic internal mode that is easily excited by 
different mechanisms (such as interaction with inhomogeneities,
\cite{internal,internal2}). Therefore, one would expect that for the $\phi^4$
equation the effect of a perturbation of a given length is more dramatic 
than for the sG model, as indeed is the case: See Fig.\ \ref{fig:lengthscale} 
for a comparison, showing that the kink is trapped for a wider range 
of values of $k$ in the $\phi^4$ equation.  
Remarkably, the present and previous results using this two 
collective coordinate approach, and particularly their interpretation 
in terms of phonons, suggest that this technique could be something 
like a `second order collective coordinate perturbation theory', 
the width degree of freedom playing the role of second order term.
We believe it is appealing to explore this possibility from a more
formal viewpoint; if this idea is correct, then one could think of 
a scheme for adding in an standard way as many collective coordinates
as needed to achieve the required accuracy. Progress in that direction
would provide the necessary mathematical grounds for this fruitful 
approximate technique.  

Finally, some comments are in order regarding the applicability of our 
results. We believe that the collective coordinate approach may also 
explain the length scale competition for breathers, that so far lacks 
any explanation. For this problem, the approach would likely involve 
the breather center and its frequency, as this magnitude controls the 
kink width when the kink is at rest. Of course, such an {\em Ansatz}
would only be valid for the breather initially at rest, and the description
of the dynamical problem would be more involved, needing perhaps more 
collective coordinates (such as an independent width). If this approach
succeeds, one can then extend it to other breather like excitations, 
such as nonlinear Schr\"odinger solitons or intrinsic localized modes. 
Work along these lines is in progress. On the other hand, it would be 
very important to have an experimental setup where all these conclusions
could be tested in the real world. A modified Josephson junction device
has been proposed recently \cite{luis2} where the role of the kink 
width is crucial in determining the performance and dynamical characteristics.
We believe that a straightforward modification of that design would 
permit an experimental verification of our results. We hope that this 
paper stimulates research in that direction.

\section*{Acknowledgments}

We thank Niurka R.\ Quintero and El\'\i as Zamora-Sillero for discussions
on the lagrangian formalism introduced in \cite{elias}. 
This work has been
supported by the Ministerio de Ciencia y Tecnolog\'\i a of Spain
through grant BFM2003-07749-C05-01. S.C. is supported by
a fellowship from the Consejer\'\i a de Edu\-ca\-ci\'on de la
Comunidad Aut\'onoma de Madrid and the Fondo Social Europeo.

%%%%% References %%%%%

\end{document}